\documentclass{pasj00}
\begin{document}
\SetRunningHead{Author(s) in page-head}{Running Head}
\Received{2007/0X/XX}%{yyyy/mm/dd}
\Accepted{200X/XX/XX}%{yyyy/mm/dd}

\title{The first determination of the actinide Th abundance for a red
giant of the Ursa Minor dwarf galaxy\thanks{Based on data collected at
Subaru Telescope, which is operated by the National Astronomical
Observatory of Japan. }}

 \author{%
  Wako \textsc{Aoki},\altaffilmark{1,2}
  Satoshi \textsc{Honda},\altaffilmark{1}
  Kozo \textsc{Sadakane},\altaffilmark{3} \\
   and
  Nobuo \textsc{Arimoto},\altaffilmark{1,2} 
}

 \altaffiltext{1}{National Astronomical Observatory, Mitaka, Tokyo 181-8588}
 \email{aoki.wako@nao.ac.jp, honda@optik.mtk.nao.ac.jp, arimoto@optik.mtk.nao.ac.jp}

 \altaffiltext{2}{Department of Astronomy, Graduate University of Advanced
  Studies, Mitaka, Tokyo 181-8588}

 \altaffiltext{3}{Astronomical Institute, Osaka Kyoiku University, Asahigaoka, Kashiwara, Osaka 582-8582}
\email{sadakane@cc.osaka-kyoiku.ac.jp}


%Do NOT move this preamble from here!

\maketitle

\begin{abstract}

The Thorium abundance for the red giant COS~82 in the Ursa Minor dwarf
spheroidal galaxy is determined based on a high resolution
spectrum. This is the first detection of actinides in an extra
Galactic object. A detailed abundance pattern is determined for 12
other neutron-capture elements from the atomic number 39 to 68. These
elements are significantly over-abundant with respect to other metals
like Fe ($>$ 1~dex) and their abundance pattern agrees well with
those of the r-process-enhanced, very metal-poor stars known in the
Galactic halo, while the metallicity of this object ([Fe/H]$\sim
-1.5$) is much higher than these field stars ([Fe/H] $\sim -3.0$).
The results indicate that the mechanism and the astrophysical site
that are responsible for neutron-capture elements in COS~82 is similar
to that for field r-process-enhanced stars, while the condition of low
mass star formation is quite different. An estimate of the age of this
object based on the Th abundance ratio is discussed.

\end{abstract}

\section{Introduction}\label{sec:intro}

The rapid process (r-process) is known to be responsible for about
half of the abundances of elements heavier than Fe in solar-system
material. Since the reaction occurs through very neutron-rich,
unstable nuclei, the understanding of the process is still a
challenging issue in nuclear physics. The astrophysical sites of this
process are still unclear, though numerical simulations have been made
for many possibilities (e.g. Wanajo \& Ishimaru 2006).

An important progress in this field in the past decade is the
discoveries of very metal-poor stars having large over-abundances of
r-process elements. Such objects are expected to record the yields of
nucleosynthetic events in the early Galaxy including the r-process.
Several objects having very large excesses of r-process elements
(e.g. Eu) with respect to other metals ([Eu/Fe]$>1$) are known in the
Galactic halo \citep{sneden03, hill02, christlieb04, barklem05}, and
are called as 'r-II' stars \citep{beers05}. A remarkable result for
these stars is that the abundance patterns of neutron-capture elements
are very similar. Objects having such large excesses of r-process
elements are only found at the metallicity around [Fe/H]$=-3.0$, and
no object having [Eu/Fe]$\gtrsim+1.0$ is known in the range of
[Fe/H]$>-2.5$ in the Galactic halo, though the reason for such
a metallicity distribution is still unknown.

However, the situation is quite different in some dwarf spheroidal
galaxies (dSph), in which high abundance ratios of r-process elements
are found even in stars with relatively high metallicity. The chemical
abundance studies for individual stars in the local group dSph made
large progresses in the past several years, thanks to high resolution
spectrographs mounted at 8--10~m telescopes (e.g. Shetrone et
al. 2001, 2003; Venn et al. 2004). One surprising result derived by
these abundance studies is the existence of stars showing extremely
large enhancements of heavy neutron-capture elements with relatively
high metallicity. The most remarkable star is the red giant COS~82 in
the Ursa Minor dSph. High resolution spectroscopy of this object was
first obtained by Shetrone et al. (2001; in their
paper, this star is referred to as '199'), who reported that this star
is moderately metal-poor ([Fe/H]$\sim -1.5$) but shows large
enhancement of heavy neutron-capture elements ([Eu/Fe]=1.49). Our
previous observing obtained a red spectrum of this
star with higher quality, and has successfully detected some other
heavy elements (e.g. Dy, Er), confirming that the abundance pattern of
elements heavier than Ba almost completely agrees with the r-process
abundance pattern estimated from solar-system material (Sadakane et
al. 2004).  This result is surprising, because no moderately
metal-poor star ([Fe/H]$\gtrsim-2$) in our Galaxy is known to date to
have such large excesses of r-process elements. Tsujimoto \& Shigeyama
(2002) applied their supernova-driven star formation model and
proposed that this phenomenon can be explained by assuming a large
velocity dispersion of the gas from which low mass stars like COS~82
formed.

In this Letter, we report the detection of the Th absorption line at
5989~{\AA} for this object by the re-analysis of our high resolution
spectrum obtained by \citet{sadakane04}. The actinide Th has an
isotope whose half life is 14.05 Gyr, and is produced only by the
r-process. The production of actinides is very sensitive to the
environment of the sites (e.g. entropy, timescale of the explosion)
according to model calculations for the r-process. Therefore, the Th
abundance can be a very strong constraint on the study of the origin
of the neutron-capture elements in this object. While Th abundance has
been determined for more than ten field metal-poor stars and globular
cluster stars, our measurement for COS~82 is the first determination
of an actinide abundance for an extra Galacitic object. We also
discuss the possibility to apply the Th abundance ratio to estimate
the age of this object.

\section{Observation, Analysis and Results}

A high resolution spectrum of COS~82 was obtained by
\citet{sadakane04} with the Subaru Telescope High Dispersion
Spectrograph (HDS, Noguchi et al. 2002). The spectrum covers
4400--7100~{\AA} with a resolving power of 40,000.  The
signal-to-noise ratio per 1.8~km$^{-1}$ pixel is 60 at 5900~{\AA}.
Standard data reduction procedures are carried out with the IRAF
echelle package\footnote{IRAF is distributed by the National Optical
  Astronomy Observatories, which is operated by the Association of
  Universities for Research in Astronomy, Inc. under cooperative
  agreement with the National Science Foundation.} as described by
\citet{aoki05}. Equivalent widths for isolated absorption lines are
measured by fitting gaussian profiles.

A standard abundance analysis was made for the measured equivalent
widths using the model atmospheres of \citet{kurucz93}. The analysis
was made primarily based on the line list of \citet{sadakane04}. We
added some weak Fe {\small I} lines, which enable one to determine the
micro-turbulence more accurately.  The effective temperature ($T_{\rm
eff}$) of 4300~K, determined by \citet{sadakane04}, was adopted in the
analysis. The micro-turbulence ($v_{\rm turb}$) and gravity ($g$) are
determined so that the derived Fe abundance is not dependent on the
strengths of Fe {\small I} lines nor on the ionization stage,
respectively. The results ($\log g=0.6$~dex and $v_{\rm
turb}=1.7$~km~s$^{-1}$) agree with those of \citet{sadakane04} within
the estimated errors. The metallicity  is assumed to be the
Fe abundance ([Fe/H]). The Fe abundance derived by the present
analysis is [Fe/H]$=-1.42$, adopting the solar abundance of
$\log\epsilon$(Fe) = 7.45 \citep{asplund05}. The results of the abundance
analysis are given in table \ref{tab:res}.

The Th abundance is determined from the Th {\small II} 5989~{\AA}
line. The oscillator strengths of Th lines, including the 5989~{\AA}
line, are obtained by \citet{nilsson02}.  This line was investigated by
\citet{yushchenko05} for the field metal-poor star HD~221170. The line
was measured for the same object, using a higher quality spectrum, by
\citet{ivans06}, who reported that the abundance determined from this
line agrees with those from other Th {\small II} lines. We performed
an analysis for HD~221170 using the spectrum obtained with the Subaru
Telescope and the model atmosphere \citep{kurucz93} with the 
parameters adopted by \citet{ivans06}, and confirmed that the
derived Th abundance agrees very well with their result. We note that
the partition function of the Th {\small II} was calculated using the
energy levels listed by \citet{bw92}.

\begin{figure}
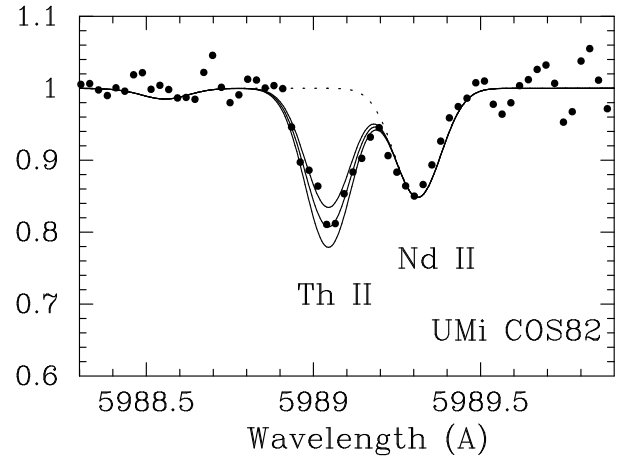

  \begin{center}
    \FigureFile(80mm,150mm){fig1.ps}
  \end{center} 
\caption{
Comparisons of synthetic spectra for the region including the Th II
5989~{\AA} line with the observed spectrum. The Th abundances assumed
in the calculations are $\log\epsilon$(Th)$= -0.15, -0.25, -0.35$
(solid lines), and $-\infty$ (dotted line). }\label{fig:sp}
\end{figure}

Though telluric absorption lines, probably due to water vapor, exist
in this wavelength range, we found that the Th {\small II} 5989~{\AA}
line of COS~82 is not distinctively affected by them.
Figure~\ref{fig:sp} shows the observed spectrum along with the
synthetic ones for three Th abundances. The data for spectral lines
other than the Th {\small II} line were adopted from the list of
\citet{kurucz95}. The wavelength of the observed spectrum is
determined from 137 Fe {\small I} lines in the spectrum. We identified
the absorption feature at 5989.3~{\AA} as a Nd {\small II} line.
Although a Nd {\small II} line at 5989.378~{\AA} with the lower
excitation potential of 6005.270 cm$^{-1}$ is listed by
\citet{kurucz95}, J. E. Lawler (private communication) suggests that
the line should be the transition from 19758.540 to 3066.755
cm$^{-1}$ at $\lambda_{\rm air} =$ 5989.312~{\AA}. Since the transition
probability of this line is unknown, we assume it to be $\log
gf=-2.05$. This $gf$-value and the Nd abundance determined from other
spectral regions well reproduce the strength of the line for COS~82 as
well as for HD~221170.

The Th abundance is determined by $\chi^{2}$-fitting of synthetic
spectra to the observed one from 5988.9 to 5989.2~{\AA}, which is not
severely affected by the 5989.3~{\AA} feature. The derived abundance
is $\log \epsilon$(Th)$=-0.25$. The 2$\sigma$ level of the fitting
error is 0.07~dex. The continuum level is estimated for the spectrum
around the absorption features. The uncertainty of the continuum
placement ($\sim 0.5$\%) results in an possible error of 0.03~dex in
the Th abundance. We estimate the line broadening for relatively clean
lines in the spectrum, and found that the line widths are primarily
determined by the instrumental broadening, and no significant
broadening by macro-turbulence nor rotation is detected. The possible
errors in the Th abundance due to the adopted broadening parameter
($\pm 0.5$~km~s$^{-1}$) is 0.03~dex.

The abundances of other elements are determined using the line list of
\citet{sadakane04} with some modifications. The transition
probabilities of Nd {\small II}, Sm {\small II} and Gd {\small II} are
updated, adding several new lines to the list, using the data of
\citet{hartog03}, \citet{lawler06} and \citet{hartog06}, respectively.
The effects of hyperfine splitting for Ba, La, Eu, and Pr are included
using the line data obtained by \citet{mcwilliam98},
\citet{lawler01a}, \citet{lawler01b}, and \citet{ginibre89},
respectively. We also measure the O abundance from the [O {\small I}]
6300.3~{\AA} line. The results are listed in table~\ref{tab:res}. An
upper limit of the Os abundance is derived from the Os {\small I}
5584~{\AA} line. Estimates of upper limits for other neutron-capture
elements are also attempted, but no meaningful result is derived from
our spectrum. The abundances of Fe and most neutron-capture
elements derived by the present analysis agrees with the results of
\citet{sadakane04} within the errors (see below). Exceptions are Eu
and Gd abundances, for which the differences between the two
measurements are about 0.25~dex. This would be due to the difference
of the spectral features and line data adopted in the analysis. The
abundances of Fe and some neutron-capture elements determined by
\citet{shetrone01} agree well with our results, while their Ce and Sm
abundances are more than 0.3~dex higher than ours. We suspect that
their results are uncertain because of the low S/N ratio in the blue
region of their spectrum where Ce and Sm lines used in their analysis
exist.

The abundance uncertainties for elements for which more than 10 lines
are measured are estimated by $\sigma N^{-1/2}$, where $\sigma$ is the
standard deviation of the abundances from individual lines and $N$ is
the number of lines used. A typical uncertainty of the equivalent
width measurement from our spectrum is 6~m{\AA}.  We performed
abundance analyses for Fe {\small I} lines with equivalent widths from
30 to 200~m{\AA}, changing the equivalent widths by 6~m{\AA}. The
effects of the change of equivalent widths on the derived abundances
are 0.15--0.20~dex, which agree with the $\sigma$ (0.18~dex) for Fe
{\small I} ($\sigma_{\rm Fe I}$). For elements whose abundances are
determined from less than 10 lines, the random error is estimated by
$\sigma_{\rm Fe I}N^{-1/2}$. Errors due to the uncertainty of
atmospheric parameters are estimated by calculating abundances
changing atmospheric parameters by $\Delta(T_{\rm eff})$=150~K,
$\Delta(\log g)$=0.3~dex, $\Delta$([Fe/H])=0.2~dex, and $\Delta(v_{\rm
turb})=0.3$~km~s$^{-1}$. These errors are added, in quadrature, to the
random errors estimated above, and are given in table~\ref{tab:res}
($\sigma_{\rm total}$). We found, however, that the abundances of the
neutron-capture elements determined from the ionized species
systematically scales by changes of the adopted atmospheric
parameters. For the discussion based on the abundance pattern, we
calculate the average of the abundance changes for heavy
neutron-capture elements (La--Er) by changing the atmospheric
parameters, and adopt the deviation from this systematic abundance
change for individual elements as the errors due to the uncertainties
of atmospheric parameters. This results in quite small errors ($\leq
0.05$~dex). These errors and the random errors are added in
quadrature, and are also given in table~\ref{tab:res} as $\sigma_{\rm
n-cap}$.

%### Figure 1

%The dashed line shows a
%calculation for the Nd {\small II} line listed by \citet{kurucz95},
%while others are calculated shifting its wavelength by $-0.06$~{\AA}
(see text).

%\clearpage

%### Figure 1
\begin{figure}
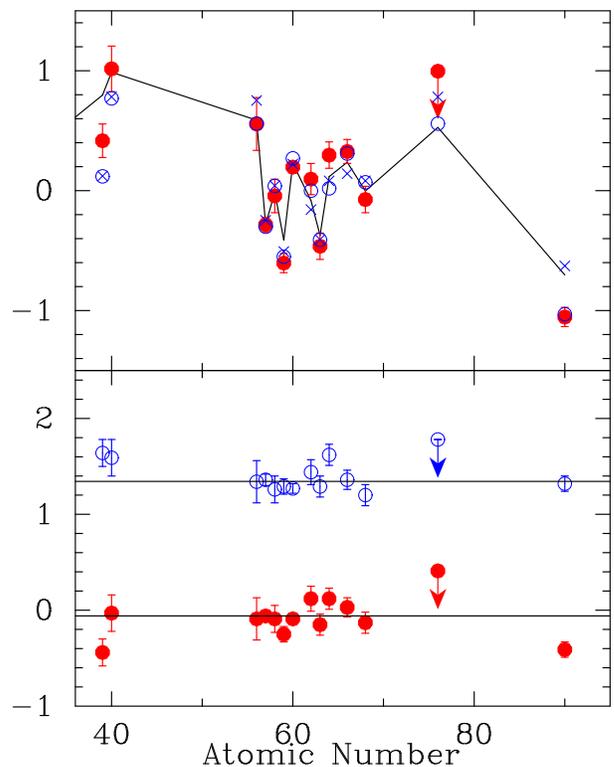

  \begin{center}
    \FigureFile(80mm,150mm){fig2.ps}
  \end{center} 
\caption{
Upper panel: Scaled abundances of neutron-capture elements for COS~82
(filled circles), CS~22892--052 (open circles), CS~31082--001
(crosses), and the solar-system r-process component (solid line). The
abundances are normalized to the average of the abundances of the 10 elements
from Ba to Er in the logarithmic scale. Lower panel: Abundance
differences in the logarithmic scale between COS~82 and CS~22892--052
(open circles) and that between COS~82 and the solar-system r-process
component (filled circles). Solid lines indicate the averages of the
abundance differences for Ba--Er.}\label{fig:pattern}
\end{figure}

\begin{longtable}{lrrrrrr}
\caption{Elemental Abundances Results}\label{tab:res}
\hline
\hline
species & $\log\epsilon_{\rm sun}$ & $\log\epsilon$& [X/Fe] & $N$ & $\sigma_{\rm totol}$ & $\sigma_{\rm n-cap}$ \\
\endhead
\hline
Fe I  & 7.45 & 6.03 &  ...  & 137 & 0.27 & ... \\
Fe II & 7.45 & 6.02 &  ...  &  13 & 0.18 & ... \\
O I   & 8.70 & 7.45 &  0.17 &   1 & 0.2  & ... \\
Y II  & 2.21 & 1.22 &  0.42 &   6 & 0.23 & 0.14 \\
Zr II & 2.59 & 1.82 &  0.65 &   1 & 0.24 & 0.19 \\
Ba II & 2.17 & 1.36 &  0.60 &   3 & 0.30 & 0.21 \\
La II & 1.13 & 0.52 &  0.81 &   8 & 0.16 & 0.06 \\
Ce II & 1.63 & 0.76 &  0.55 &   4 & 0.20 & 0.14 \\
Pr II & 0.80 & 0.20 &  0.82 &   7 & 0.19 & 0.11 \\
Nd II & 1.45 & 1.00 &  0.97 &  41 & 0.17 & 0.05 \\
Sm II & 0.98 & 0.90 &  1.34 &   2 & 0.21 & 0.13 \\
Eu II & 0.52 & 0.34 &  1.24 &   3 & 0.17 & 0.11 \\
Gd II & 1.09 & 1.10 &  1.43 &   3 & 0.16 & 0.12 \\
Dy II & 1.17 & 1.13 &  1.38 &   3 & 0.17 & 0.10 \\
Er II & 0.97 & 0.73 &  1.18 &   3 & 0.17 & 0.11 \\
Th II & 0.09 &-0.25 &  1.08 &   1 & 0.15 & 0.08 \\
\noalign{\smallskip}
\hline
\end{longtable}

\section{Discussion and concluding remarks}

The upper panel of figure~\ref{fig:pattern} shows the abundance patterns
of neutron-capture elements for COS~82, the solar-system r-process
component \citep{simmerer04}, and the two r-process-enhanced field
stars CS~22892--052 \citep{sneden03} and CS~31082--001
\citep{hill02}. The abundances are normalized to the average of Ba--Er
ones. As reported by \citet{sadakane04}, the overall abundance pattern
of heavy neutron-capture elements (Ba--Er) in COS~82 agrees well with
the r-process abundance pattern. In the lower panel, the abundance
differences between COS~82 and the solar-system r-process component, as
well as that between COS~82 and CS~22892--052, are shown. The Th
abundance of COS~82 is lower than the normalized value of the
solar-system r-process component, while that agrees very well with the
value of CS~22892--052. This point is discussed below in more
detail. The abundance of the light neutron-capture element Y in COS~82
is also lower than the normalized value in the solar-system r-process
component as found in field r-process-enhanced stars.

%. A similar trend is known in the r-process-enhanced field
%stars including CS~22892--052, and it is interpreted that there is
%another component that contribute to the light neutron-capture
%elements in the solar-system r-process component than the component
%represented by r-process-enhanced stars.

%The normalized Y and Zr abundances of COS~82 are slightly higher than
%those of CS~22892-052, possibly indicating that another process also
%contributed to these elements to COS~82.

Although such small deviation exists, an important result derived from
the comparison is that the abundance pattern of neutron-capture
elements, including Th, in COS~82 agrees with the pure r-process
abundance pattern. According to model calculations for the r-process,
the actinide abundances with respect to other neutron-capture elements
are very sensitive to the model parameters (e.g. electron fraction,
entropy; Wanajo et al. 2002). The agreement of the Th abundance ratio
to other neutron-capture elements in COS~82 with that of CS~22892--052
indicates that the mechanism and astrophysical sites that are
responsible for the heavy elements in COS~82 is common with those for
the r-process-enhanced field stars. Recent abundance studies for
r-process-enhanced stars suggest that the r-process occurs under a
quite limited condition (e.g. supernova explosions whose progenitors
have mass in a very narrow range). Our result shows that this is also
applied to stars even in dSph. 

The metallicity of COS~82 is more than one order of magnitude higher
than those in the field r-II stars ([Fe/H]$\sim -3$). The chemical
composition of field stars with such metallicity is believed to be
affected by a number of nucleosynthesis events.  However, the
metallicity, as well as the amount of neutron-capture elements, should
be related to how the yields from a nucleosynthesis event are captured
by the clouds from which next generation, low mass stars formed, and
the neutron-capture elements of COS~82 might be provided by a single
event. For instance, \citet{tsujimoto02} proposed that the large
excesses of neutron-capture elements in a few objects in dSph,
including COS~82, are the result of a larger velocity dispersion of
the gas from which low mass stars like COS~82 formed in dSph than in
the Galactic halo.  That is, the yields of one supernova are recorded
in the next generations of stars with smaller mixing with
metal-deficient (or metal-free) gas in dSph.
  
Finally, we discuss an estimate of the age of COS~82 from the abundance
ratio between Th and other stable elements, which could be an
independent calibration for the age measurements for this galaxy. If
the age of this star is shown to be quite long ($>10$~Gyr), the star
formation of this dSph should be completed with a short time scale.
Th has a half life of 14.05 Gyr, and the age of the object (more
strictly, the duration from the r-process event) is estimated from the
Th abundance ratio to other stable elements, if its initial abundance
ratio is given. Here the average of abundances of the nine elements
from La to Er is used as the reference \footnote{Ba is excluded from
the reference elements, because its abundance error is relatively
large due to the strengths of the lines used in the analysis. The
conclusion in this paragraph is, however, unchanged if this element is
included in the calculation.} (E$_{\rm stable}$).  The Th/E$_{\rm
stable}$ ratio of COS~82 agrees very well with that of CS~22892--052,
while that is 0.3~dex lower than the value in the r-process component in
solar-system material (figure~\ref{fig:pattern}). 

Unfortunately, the abundance ratio of Th/E$_{\rm stable}$ produced by
the r-process is not constant. Indeed, the actinides including Th are
significantly enhanced, with respect to other neutron-capture
elements, in the r-II star CS~31082--001, compared to that of the
solar-system r-process component \citep{hill02}. However, studies of
Th abundances for a number of r-process enhanced field stars show that
the abundance ratios distribute in a rather narrow range ($-0.6\leq
$Th/E$_{\rm stable} \leq -0.1$), and CS~22892--052 is one of the
objects that have the lowest ratio. Assuming that the initial
abundance ratio of Th/E$_{\rm stable}$ for COS~82 is within this
range, and that the field halo stars including CS~22892--052 are quite
old (their ages are around 12--13 Gyr), COS~82 is also an old object.
(If a higher value of the initial Th/E$_{\rm stable}$ ratio is
assumed, the derived age becomes longer).  

The error ($\sigma_{\rm n-cap}$) in the determination of the Th
abundance ratio is 0.08~dex, while the scatter of the abundances of
the above nine stable elements around those of the solar-system
r-process component, as well as of CS~22892--052, is 0.11~dex. We
estimate the error of the Th/E$_{\rm stable}$ to be 0.13~dex, which
results in the uncertainty of the age estimate to be 6 Gyr. Hence,
though the object is suggested to be as old as field halo stars, the
age estimate using Th chronometer gives only a weak constraint. A
measurement of U/Th ratio, which is a more sensitive chronometer, for
this object will enable one to estimate the age with a much smaller
error ($\lesssim$ 2 Gyr). \\

We are grateful to Dr. J. E. Lawler for providing useful Nd {\small
  II} line data. 

%\input{refe}

%\clearpage
%%%%%%%%%%%%%%%%%%%%%%%%%%%%%%%%%%%%%%%

%\input{table}

%Na I  & 6.17 & 3.82 & -0.93 &   2 & 0.41 &  \\
%Mg I  & 7.53 & 6.31 &  0.20 &   1 & 0.31 &  \\
%Si I  & 7.51 & 6.13 &  0.04 &   1 & 0.18 &  \\
%Ca I  & 6.31 & 4.80 & -0.09 &  22 & 0.25 &  \\
%Sc II & 3.05 & 1.19 & -0.45 &   6 & 0.15 &  \\
%Ti I  & 4.90 & 3.15 & -0.33 &  18 & 0.38 &  \\
%Ti II & 4.90 & 3.10 & -0.39 &   7 & 0.15 &  \\
%V I   & 4.02 & 2.21 & -0.40 &   9 & 0.32 &  \\
%Cr I  & 5.64 & 4.01 & -0.21 &   7 & 0.38 &  \\
%Mn I  & 5.39 & 3.63 & -0.34 &   9 & 0.29 &  \\
%Co I  & 4.92 & 2.98 & -0.52 &   2 & 0.29 &  \\
%Ni I  & 6.23 & 4.40 & -0.41 &  15 & 0.23 &  \\
%Zn I  & 4.60 & 2.87 & -0.31 &   1 & 0.22 &  \\

\end{document}